\documentclass[twocolumn,aps,prl,letterpaper,nofootinbib,raggedbottom,nobalancelastpage]{revtex4}
\usepackage{graphicx,bm}
\usepackage{amsmath}
\usepackage{color}

\begin{document}
\title{Generation of ``triggered single photons'' from a coherently-pumped quantum dot}
\date{\today}
\author{P. K. Pathak and S. Hughes}
\address{Department of Physics,
Queen's University, Kingston, ON K7L 3N6, Canada}

\maketitle

\textbf{A deterministic ``on demand'' source of single photons is a basic building block for linear quantum
computation~\cite{linear}, quantum cryptography~\cite{crypto}, quantum teleportation~\cite{teleport}, and quantum networks~\cite{network}. In all these applications, quantum interference between two single-photon pulses on a symmetric beam splitter has been exploited~\cite{review}, which imposes stringent requirement for the implemented single photons to be indistinguishable in all degrees of freedom, including their frequencies, spectral widths, pulse shapes, and polarizations. To generate single photons one requires a pumping mechanism to excite a ``two-level emitter'' and an efficient channeling of the subsequently emitted photons. The efficiency of the source can be enhanced by coupling the emitter to the waveguide~\cite{manga} or a microcavity mode~\cite{cklaw}. However, hitherto the solid-state single photon sources realized by using a quantum dot coupled with a microcavity rely on incoherent pumping of excitons, which leads to problems with timing jitter~\cite{kiraz}, leading to a trade off between efficiency and indistinguishability.  Here we introduce
  a means to realize a highly efficient solid state source of indistinguishable single photons using cavity-assisted adiabatic Raman passage in a single quantum dot
 - cavity system. We demonstrate pulse-triggered single photons with  100\% efficiency and $>90$\% indistinguishability using currently available experimental parameters.}

 Single photon sources have been
 been realized in  a plethora of different quantum systems,
  including
  {\em inter alia}: single atom trapped in optical cavity~\cite{rempeold}, trapped ions~\cite{ion}, molecules \cite{molecule}, and quantum dots (QD)~\cite{incohrent,indis}. Semiconductor QDs
  have discrete energy levels similar to those in atoms and ions due to strong quantum confinement of the electron-hole pairs, and they can be embedded or grown with high precision in different semiconductor microcavities at desired spatial
  positions. The strong coupling regime, whereby the emitted photon become entangled with the cavity mode, has also been demonstrated~\cite{photocavity1,photocavity2}. Consequently, QDs provide an excellent opportunity for realizing quantum optical phenomena in solid-state systems; in  particular, for the purpose of quantum information processing, they provide outstanding potential advantages of integrability and scalability.
However,
QD-cavity-enabled  single photon sources
rely on {\em incoherent} pumping of excitons or electron-hole pairs~\cite{laussy}. Through incoherent pumping, the QD is excited in a quantum state far above from the desired exciton state, which relaxes quickly to the desired exciton state by phonon interactions. Thus the excited state has time uncertainty, resulting in so-called {\em timing jitter}. This is a major problem as
 the incoherent pumping of a QD cannot provide indistinguishable photons and high efficiency simultaneously~\cite{kiraz}. While there have been great achievements in improving the semiconductor cavity coupling and output efficiency of the emitted photons, there has been little work to address coherent on-demand
 loading.
 The coherent manipulation of energy levels in QDs has been a challenging task because of unavoidable photon scattering from the excitation optical source. In the last year, the coherent manipulation of exciton states in QDs has been demonstrated 
 in a few remarkable experiments~\cite{xu,press,autler,muller}, using, for example, two pulse excitation and the Autler-Townes splitting of dressed-exciton states. Recently, Flagg {\em et al.}~\cite{flagg} and Ates {\em et al.}~\cite{ates} have also demonstrated resonance fluorescence from a QD coupled to a semiconductor microcavity. In their experiment, the QD has been coherently driven by the external laser field and the emitted photons have been collected through the cavity mode, where the latter is geometrically separated from the excitation field.

\begin{figure}[t!]
\centering
\includegraphics[height=3.2in,width=3.2in]{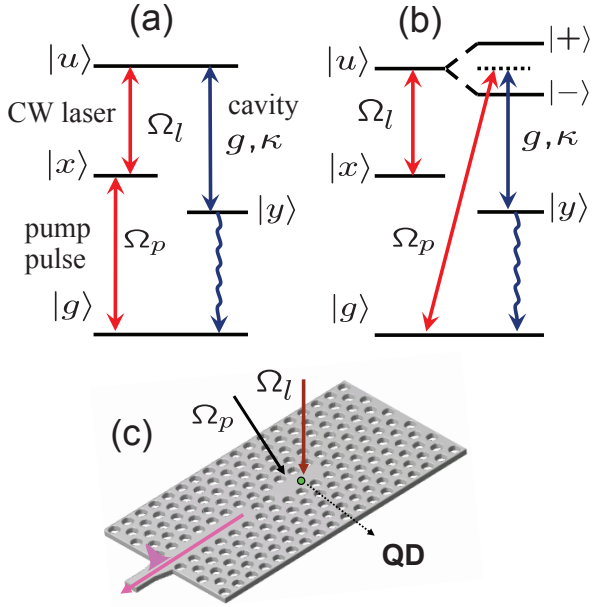}
\vspace{-0.1cm}
\caption{(Color online)
(a) Schematic of the single photon source using the Raman flip process in a QD, using a CW-control laser and a pump pulse laser, and (b) the equivalent scheme in a dressed state picture. (c)
Schematic of a planar photonic crystal cavity system as an example semiconductor
cavity system that would also emit the single photons on-chip.} \label{fig1}
\end{figure}

In this work, we
introduce
 an original
 optical
 excitation scheme
that
exploits the stimulated Raman adiabatic passage (STIRAP) in a QD embedded in a semiconductor microcavity
to coherently generate single photons. We show that the two-photon coherent Raman transitions in QDs~\cite{ramandot} can be realized through Autlet-Townes doublet.
We consider a QD embedded in a semiconductor microcavity, where
energy levels of the system is shown in Fig.~\ref{fig1}(a). The transitions from the biexciton state $|u\rangle$ to the exciton states $|x\rangle$  and $|y\rangle$ are coupled by a $x$-polarized laser field with Rabi frequency $\Omega_l$, and a $y$-polarized cavity mode with vacuum Rabi coupling $g$, respectively. Because of the large biexciton binding energy in QDs, the laser field and the cavity mode effectively remain uncoupled with the transitions from the ground state $|g\rangle$ to the exciton states. The QD is optically pumped from its ground state $|g\rangle$ to the exciton state $|x\rangle$ by applying an $x$-polarized pump pulse which has Rabi frequency $\Omega_p(t)$.
Somewhat remarkably, this scheme actually benefits from the natural anisotropic-exchange splitting that occurs
between the $x$ and $y$ polarized excitons, which is usually a major problem
for creating entangled photon pairs~\cite{stevenson2006}.
 The Hamiltonian of the system in the rotating frame can be written as
$H=\hbar\Delta_p|x\rangle\langle x|+\hbar(\Delta_p+\Delta_l-\Delta_c)|y\rangle\langle y|+\hbar(\Delta_p+\Delta_l)|u\rangle\langle u|
-\hbar\left[\Omega_l|u\rangle\langle x|+\Omega_p(t)|x\rangle\langle g|+g|u\rangle\langle y|a+H.c.\right],
$
where $\Delta_p$, $\Delta_l$, and $\Delta_c$ are the detunings of the pump pulse, the laser field, and the cavity mode respectively. For simulating the complete evolution of the system, we follow Master equation calculations in the density matrix representation. The evolution of the QD-cavity system is given by the equation 
\begin{eqnarray}
\frac{\partial\rho}{\partial t}=-\frac{i}{\hbar}[H,\rho]-\frac{1}{2}\sum_{\mu}L_{\mu}^{\dag}L_{\mu}\rho -2L_{\mu}\rho L_{\mu}^{\dag}+\rho L_{\mu}^{\dag}L_{\mu} ,
\label{master}
\end{eqnarray}
where $L_{\mu}$ are the Lindblad operators, where
$\sqrt{\gamma_1}|x\rangle\langle u|$, $\sqrt{\gamma_1}|y\rangle\langle u|$, $\sqrt{\gamma_2}|g\rangle\langle x|$, and $\sqrt{\gamma_2}|g\rangle\langle y|$ correspond to the spontaneous decays, and $\sqrt{2\gamma_d}|u\rangle\langle u|$, $\sqrt{\gamma_d}|x\rangle\langle x|$, and $\sqrt{\gamma_d}|y\rangle\langle y|$ correspond to the dephasing of biexciton and exciton states. The emission of the single photon pulses from the cavity mode is given by the Lindblad operator $\sqrt{\kappa}a$.
Initially, the QD is
in the ground state $|g\rangle$ and the cavity mode in the vacuum state. The resonant $x$-polarized laser field is applied between the states $|x\rangle$ and $|u\rangle$. In the presence of the laser field the biexciton state $|u\rangle$ and the exciton state $|x\rangle$ form the Autler-Townes doublet, $|\pm\rangle=(|u\rangle\pm|x\rangle)/\sqrt{2}$, shown in Fig.~\ref{fig1}(b). The transitions from the states $|\pm\rangle$ to the ground state $|g\rangle$ and the exciton state $|y\rangle$ are dipole allowed, and the dipole couplings with the states $|g\rangle$ and $|y\rangle$ for each Autler-Townes state remain equal for the resonant laser field $\Omega_l$, i.e. for $\Delta_l=0$. This is a necessary requirement for the complete population transfer in STIRAP through multiple intermediate states \cite{mstirap}. Due to the
Autler-Townes splitting, the applied pump pulse and the cavity mode get detuned by $\Delta_p\pm\Omega_l/2$ and $\Delta_c\pm\Omega_l/2$, respectively. Therefore, for $\Delta_p=\Delta_c$ the pump pulse and cavity mode together satisfy the two-photon Raman resonance condition. When a slowly varying $x$-polarized pump pulse $\Omega_p$, resonant with $|g\rangle$ to $|x\rangle$ transition, is applied in the presence of laser field $\Omega_l$, the QD faithfully follows the cavity-assisted Raman adiabatic passage. In such conditions, the initial state of the cavity-QD system $|g,0\rangle$ is almost adiabatically transferred to the state $|y,1\rangle$, and the populations in the states $|x,0\rangle$ and $|u,0\rangle$ should remain negligible throughout the pump pulse (we quantify this assumption below with rigorous calculations). From the state $|y,1\rangle$, a single $y$-polarized photon is emitted from the cavity mode with emission rate $\kappa$ and the QD is left in the exciton state $|y\rangle$ which very slowly decays to the ground state $|g\rangle$ with spontaneous decay rate $\gamma_2$. After a time $\approx6/\gamma_2$, when the entire population is returned back to ground state $|g\rangle$ the next pulse is applied. In our scheme, the population is further recycled back naturally and so no recycling pulses are necessary as required in the case of trapped atoms~\cite{rempeold}. However, if one wanted to reduce the time interval between generated single photon pulses, one could use a $y$-polarized $\pi$-pulse resonant with $|y\rangle$ to $|g\rangle$ transition after the emission of the single photon. Various semiconductor cavity systems  now realize such energy configurations,
and Fig.~2(c) shows an example of a photonic crrystal system where the
cavity coupling parameters
can be controlled in a systematic way~\cite{peijunPRB2009}.

\begin{figure}[b!]
\centering
\includegraphics[height=2in,width=3in]{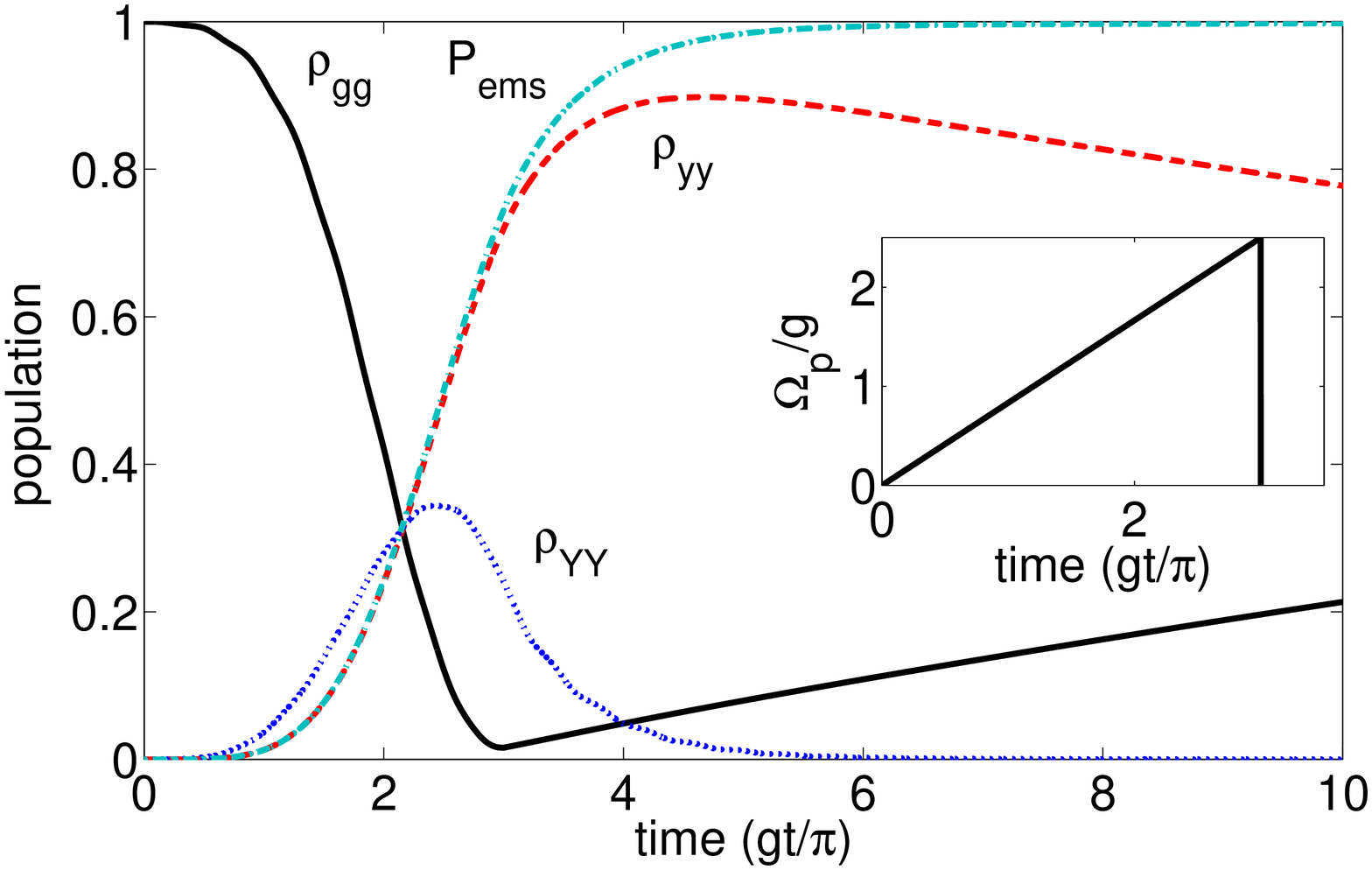}
\vspace{-0.3cm}
\caption{(Color online) The populations of the states $|g,0\rangle$ ($\rho_{gg}:$ black solid), $|y,0\rangle$
($\rho_{yy}:$ red dashed), and $|y,1\rangle$ ($\rho_{YY}:$ blue dotted) of the QD-cavity system and the emission probability, $P_{\rm ems}$ (green chain), from the cavity mode for one pump pulse using parameters $\Delta_p=\Delta_l=\Delta_c=0$, $\Omega_l/g=5$, $\gamma_1/g=\gamma_2/g=\gamma_d/g=0.01$, and $\kappa/g=0.5$. The pump pulse is chosen as sawtooth wave with maximum amplitude $\Omega_{\rm max}=2.5g$ and pulse widths $gt_p=3\pi$, which is shown in the inset.} \label{fig2}
\end{figure}

To maximize the STIRAP population transfer, we
use a sawtooth wave pulse of pulse-width given by $gt_{p}=3\pi$ with peak amplitude $2.5g$, which is applied between ground  state $|g\rangle$ to $|x\rangle$.
In Fig.~\ref{fig2}, we show the populations of the quantum states of the QD-cavity system, $\langle i|\rho|i\rangle$ for $|i\rangle=|g,0\rangle,~|y,1\rangle$, and $|y,0\rangle$, calculated using Eq.~(\ref{master}) (see Supplementary Information).
To label the various population levels we introduce the following
notation:
$\rho_{gg}$ - ground state population,
$\rho_{YY}$ - $y-$polarized exciton and a cavity photon,
$\rho_{yy}$ - $y-$polarized exciton and an emitted cavity photon;
we also define
$P_{\rm ems}(t)=\int_0^t \kappa n_c(t') dt'$ as the probability of cavity photon
emission at time $t$.
The initial ground state population ($\rho_{gg}$) decreases and reaches
a minimum during the pulse; however,
it never becomes zero because of a small population returning back in the ground state from the spontaneous decay of the state $|y,0\rangle$ ($\rho_{yy}$). The population
 in the state $|y,1\rangle$ ($\rho_{YY}$),
 reaches the maximum before the pump pulse attains its maximum,
 which then
  decays to the state
  $|y,0\rangle$
  after emitting a single photon pulse. A small population of the order of $10^{-3}$ is also generated in the state
  $|g,1\rangle$ through the spontaneous decay of the state $|y,1\rangle$.
   The populations in the state $|x,0\rangle$ and $|u,0\rangle$ remain of the order of $10^{-2}$ during the pulse, thus the evolution of the QD-cavity system efficiently follows Raman adiabatic passage. The probability of the photon emission from the cavity mode during the pulse,
  $P_{\rm emiss}$
  reaches $1$ when the population in the state $|y,1\rangle$ decays to zero (see Fig.~\ref{fig2}).

As discussed above, because of the spontaneous decay of exciton state $|y,0\rangle$, there is a small continuous flow in the population of the ground state $|g,0\rangle$ during the pump pulse. If a large population is returned back in ground state during the pump pulse it could lead to the emission of more than one photon from the cavity mode in every pump pulse. To avoid the emission of more than one photon per pulse, the pulse-width should be chosen such that $\gamma_2 t_p<0.1$, which comes from the fact that, for $\gamma_2=0$, the emission probability per pulse remains larger than $0.9$ and the contribution of the population in ground state from the spontaneous emission is $\approx\gamma_2t_p$. For larger values of peak amplitude the required pulse widths for complete population transfer are smaller, however to avoid populating state $|u,0\rangle$ and $|x,0\rangle$ the peak amplitude of the applied pump pulse should be smaller than the detunings from the upper states $|\pm\rangle$, i.e. $\Omega_p(t)<\Omega_l$. For QD embedded in a micro cavity the off-resonant exciton have the spontaneous decay rate of the order of $0.1-1\,\mu$eV~\cite{photocavity1,photocavity2} ($10^{-2} g$ if
 $g = 20\,\mu$eV), thus the condition $\gamma_2t_p<0.1$ or $t_p<4-40$\,ns can easily be satisfied in present experiments.

\begin{figure}[t]
\centering
\includegraphics[height=2in,width=3in]{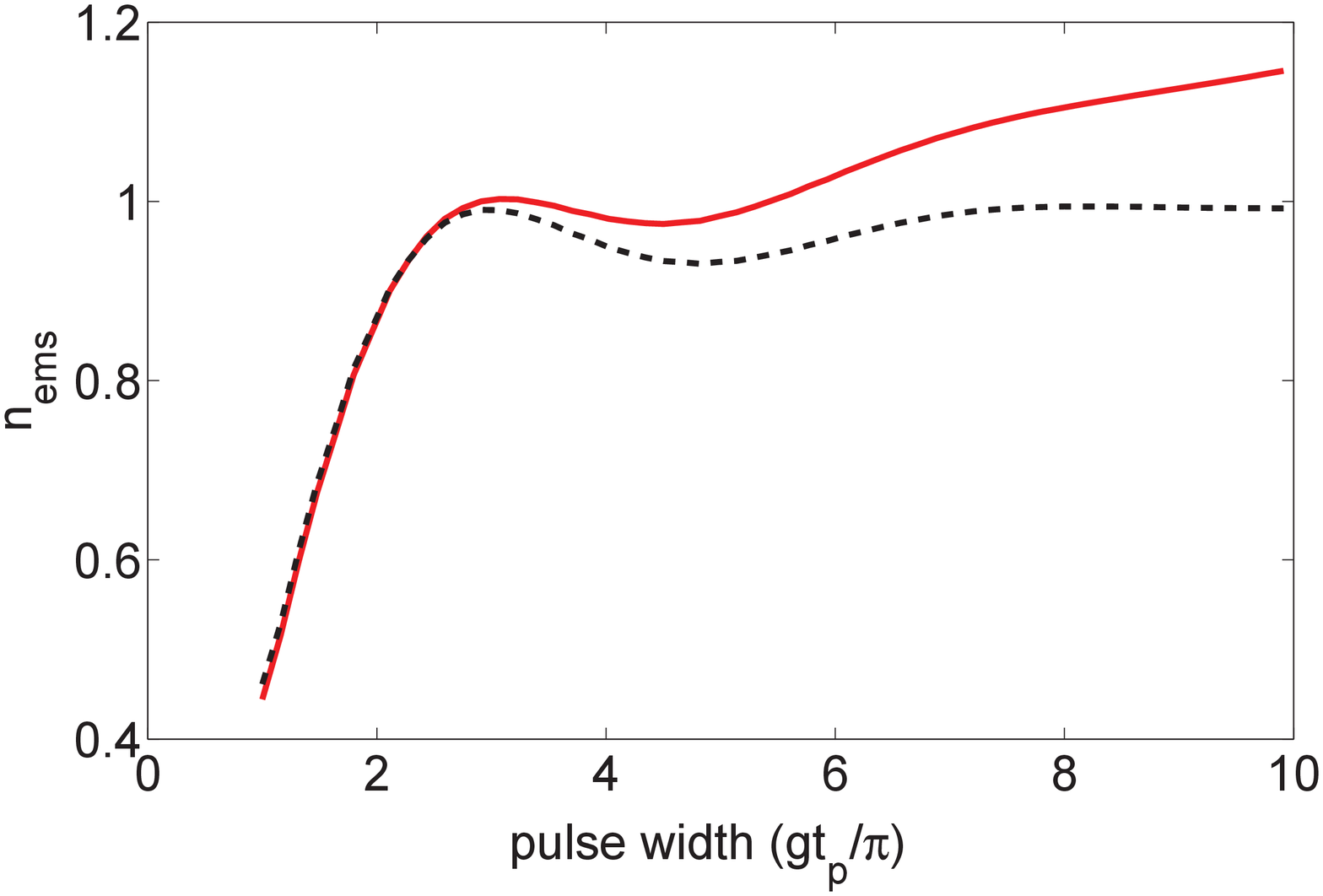}
\vspace{-0.3cm}
\caption{(Color online) Pulse width dependence
of the number of photons emitted from the cavity mode, $n_{\rm ems}$,
 for one pump pulse of sawtooth wave shape. The red solid (black dashed) curve is for $\gamma_2/g=0.01$ ($\gamma_2/g=0$). The other parameters are the same as in Fig.~\ref{fig2}.} \label{fig3}
\end{figure}

In Fig.~\ref{fig3}, we show the number of photons emitted in a pump pulse with increasing pulse widths. The lower red curve is for $\gamma_2=0$ and the upper black curve is for $\gamma_2/g=0.01$. After the complete transfer of ground state population in $|y,1\rangle$ and subsequent emission of a single photon pulse the population in state $|y,0\rangle$ becomes large and starts contributing to the population in ground state. If the pulse-width is larger than the required value for complete population transfer in STIRAP, the population transferred to ground state in spontaneous emission of $|y,0\rangle$ can contribute to the photon emission per pulse and the number of photons emitted per pulse rises on increasing pulse-width. After complete population transfer in STIRAP, for the two-photon Rabi coupling larger than the cavity mode decay rate, i.e. $g\Omega_p/\Omega_l\geq\kappa$, a small population from $|y,1\rangle$ reflects back to the state $|x,0\rangle$ and the emitted number of photons per pulse remains nearly one for a long range of pulse widths even after a small feedback in the ground state population from the spontaneous decay of the state $|y,0\rangle$. This can also provide more flexibility in choosing pump pulse-widths. However, for $g\Omega_p/\Omega_l>\kappa$, the number of photons emitted per pulse becomes larger than one on increasing pulse widths after complete population transfer in STIRAP.

\begin{figure}[t]
\centering
\includegraphics[height=2in,width=3in]{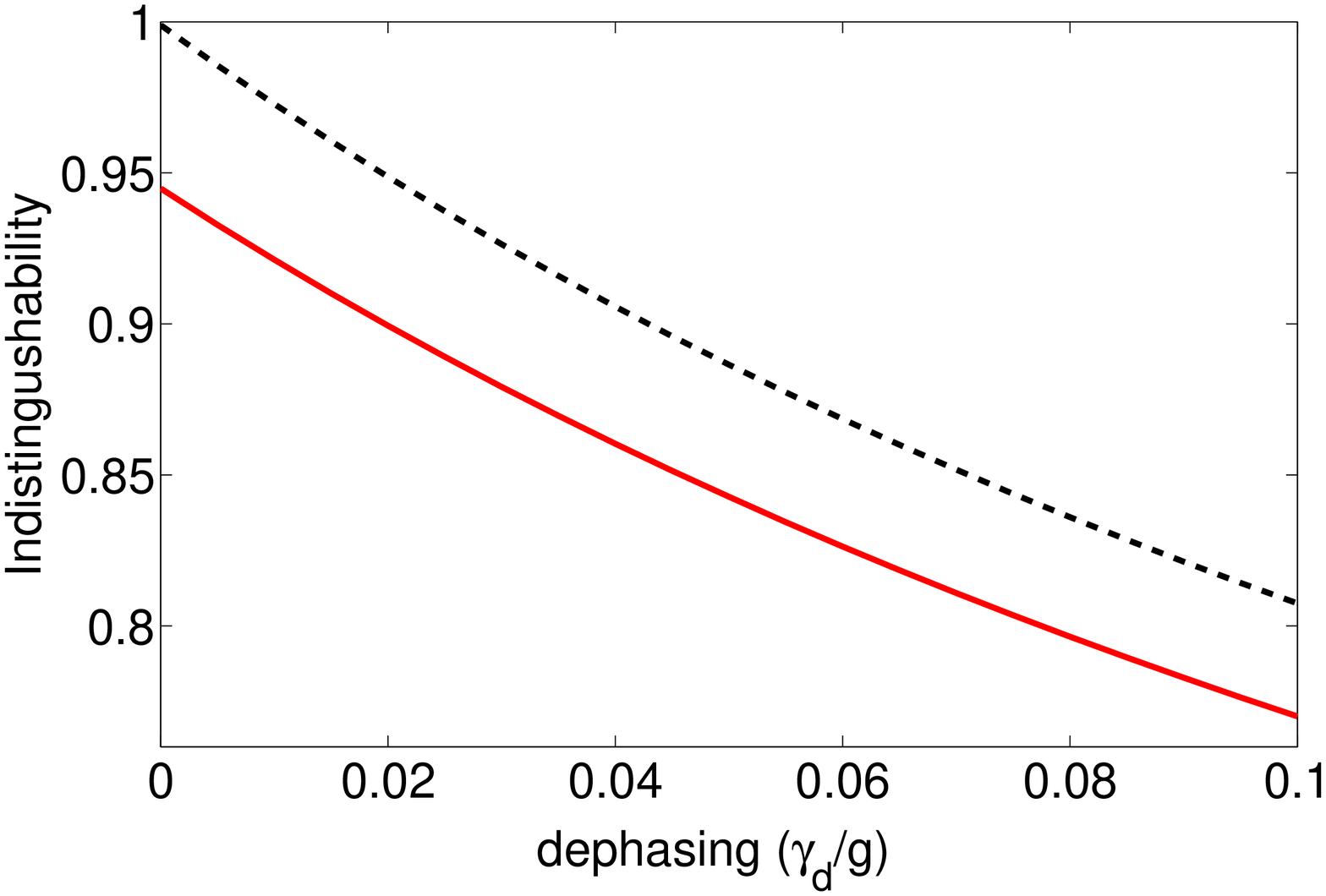}
\vspace{-0.3cm}
\caption{(Color online) The dependence of the indistinguishability of generated photon pulses on dephasing rate $\gamma_d$. The red solid (black) curve is for $\gamma_2/g=0.01$ ($\gamma_2=0.0$), the maximum pulse amplitude $2.5g$ and pulse width $gt_p=3\pi$, respectively. The other parameters are the same as in Fig.~\ref{fig2}.} \label{fig4}
\end{figure}

Next, we analyze the indistinguishability of the generated photons, which is an important aspect in the context of the applicability of the single photon source in quantum information protocols. The indistinguishability of the photons is measured by sending two consecutive generated photon pulses through a beam splitter and detecting Hong-Ou-Mandel type correlation \cite{hong}. For perfectly indistinguishable photons the probability of coincidence detection at output ports of a symmetric beam-splitter remains zero because of the Bose-Einstein statistics. In the earlier experiments~\cite{indis}, for measuring indistinguishability, the photons are allowed to pass through a Michelson interferometer or through a Mach-Zhender interferometer having path difference between two arms corresponding to produce a time delay between photons, which is equals to the time difference between two photon pulses generated from the single photon source. In such case the two photons are incident on a symmetric beam splitter at the same time. After passing through the interferometer the probability of the coincidence detection of one photon at each output port can be expressed in terms of the correlations of the cavity field operator as follows~\cite{kiraz}:
\begin{equation}
P_c=\frac{1}{2}\left[1+\frac{\int_0^T dt\int_0^{T-t}d\tau [g^{(2)}(t,\tau)-|g^{(1)}(t,\tau)|^2]}{\int_0^T dt\int_0^{T-t}\langle a^{\dagger}(t)a(t)\rangle\langle a^{\dagger}(t+\tau)a(t+\tau)\rangle}\right],
\label{pcoin}
\end{equation}
where $T$ is the time interval between two pump pulses, $g^{(1)}(t,\tau)=\langle a^{\dagger}(t)a(t+\tau)\rangle$ and $g^{(2)}(t,\tau)=\langle a^{\dagger}(t)a^{\dagger}(t+\tau)a(t+\tau)a(t)\rangle$ are the unnormalized first-order and second-order quantum correlation functions of the cavity field. The first-order correlation contains the interference effects between two photons and the second-order correlation contains the probability of generating more than one photon per pulse in the source. We calculate the different correlations required in Eq.~(\ref{pcoin}) by solving optical Bloch equations for one time correlations from Eq.~(\ref{master}) and using quantum regression theorem (see Supplementary Information). The indistinguishability of the photons is given by $1-P_c$. In Fig.\ref{fig4}, we show the dependence of the indistinguishability of the generated photons on dephasing of the biexciton and the exciton states. The indistinguishability of the photons decreases on increasing dephasing rate. The decrease in indistinguishability can be understood as a consequence of a small information about the photons gets stored in different baths attached to the emission of photon in the cavity mode. In our scheme there are phonon baths attached with the biexciton state $|u\rangle$ and exciton state $|y\rangle$ corresponding to the dephasing and thermal baths corresponding to the spontaneous emission from state $|u\rangle$ to $|y\rangle$ and $|y\rangle$ to $|g\rangle$. The spontaneous decay of exciton $|y\rangle$ also contribute to a small probability of generating more than one photon per pulse which adversely affects the indistinguishability of photons. However, the decay rate $\gamma_1$ does not affect the indistinguishability much and only a decrease of the order of $10^{-3}$ occurs by changing the $\gamma_1$ from 0 to 0.1. Thus we believe
our proposed scheme is possible using a range of present day cavity - QD systems.

Before concluding, we remark about possible experimental setups.
Our scheme could use a QD embedded in a photonic crystal cavity having single linear polarized mode and shining the driving laser and pump pulse similar to the experiment of coherent excitation of QD in PC-cavity~\cite{jelena}. Another cavity system could consist of a
 QD embedded inside a high quality micropillar cavity and using orthogonal excitation-detection techniques.  The linearly polarized excitation pulses, resonant with $|g\rangle$ to $|x\rangle$ transition, and driving laser, resonant with the $|u\rangle$ to $|x\rangle$ transition, can be implemented together similar to the $s-$shell excitation method in a recent experiment by Ates
 {\em et al.}~\cite{ates}. The linearly polarized cavity mode, coupled with $|u\rangle$ to $|y\rangle$ transition, can be realized in micro-pillars having elliptical cross-section~\cite{ellip}.

In conclusion, we have presented a new scheme  that shows unprecedented promise for generating highly efficient solid state source of indistinguishable single photons using STIRAP in a single QD-cavity system. The two-photon Raman transition in the QD can be realized through the Autler-Townes doublet generated by using a resonant laser field between biexciton and exciton states. The generated single photon pulses has a linear polarization orthogonal to the driving field, which makes it distinguishable from the background and well suited for implementation of polarization encoded qubit and pulse shaping \cite{shaping}. We have shown that in our scheme single photons can be generated with 100\% efficiency and more than 90\% indistinguishability in the Hong-Ou-Mandel type interference, using currently available experimental parameters. We have also discussed the possible experimental implementations
using present day technologies.

\section*{Acknowledgements} 
\label{sec:acknowledgements}
This work was supported by the National Sciences and Engineering Research Council of Canada,
and the Canadian Foundation for Innovation. We acknowledge discussions with R.L. Williams.


\end{document}